\documentclass[useAMS,usenatbib]{mn2e}

\usepackage{caption}
\usepackage{graphicx}
\usepackage{epstopdf}
\usepackage{amsmath}
\usepackage{array}
\usepackage{units}

\def\Kepler{\textit{Kepler}}

\title[Accretion induced variability in CVs]
{The Universal Nature of Accretion-induced Variability: The RMS-Flux Relation in an Accreting White Dwarf}
\author[S. Scaringi \textit{et al.}]
{S. Scaringi$^{1}$\thanks{E-mail: s.scaringi@astro.ru.nl}, E. K\"{o}rding$^{1}$, P. Uttley$^{2,3}$, C. Knigge$^{2}$, P.J. Groot$^{1}$, M. Still$^{4,5}$\\ 
$^{1}$Department of Astrophysics/IMAPP, Radboud University Nijmegen, P.O. Box 9010, 6500 GL Nijmegen, The Netherlands \\ 
$^{2}$Department of Physics and Astronomy, University of Southampton, Highfield, Southampton, SO17 1BJ, UK \\ 
$^{3}$Astronomical Institute ``Anton Pannekoek'', University of Amsterdam, Science Park 904, 1098XH, Amsterdam, The Netherlands \\
$^{4}$NASA Ames Research Center, Moffett Field, CA 94035, USA \\
$^{5}$Bay Area Environmental Research Institute, Inc., 560 Third St.West, Sonoma, CA 95476, USA \\
}

\begin{document} 

\date{}

\pagerange{\pageref{firstpage}--\pageref{lastpage}} \pubyear{2011}

\maketitle

\label{firstpage}

\begin{abstract}
We report the discovery of a linear relationship between the root-mean-square (rms) variability amplitude and the mean flux in the accreting white dwarf binary system MV Lyrae. Our lightcurve, obtained with the \Kepler\ satellite, spans 633 days with quasi-continuous 58.8 second cadence resolution. We show, for the first time, how this cataclysmic variable displays linear rms-flux relations similar to those observed in many other black hole binaries, neutron star binaries and Active Galactic Nuclei. The phenomenological similarity between the rms-flux relation observed here and in other X-ray binaries suggests a common physical origin for the broad-band variability, independent of source type, mass or size of the compact accretor. Furthermore, we infer the viscosity parameter, $\alpha$, and disk scale height, $H/R$, using two independent methods. In both cases, both values are found to be uncomfortably high to be accommodated by the disk instability model.

\end{abstract}


\section{Introduction}
Compact interacting binaries (CBs) are close binary systems usually consisting of a late-type star transferring material onto a compact object such as a black hole (BH), neutron star (NS) or a white dwarf (WD) via Roche-lobe overflow. With an orbital period on the order of hours, the transferred material from the donor star forms an accretion disk surrounding the compact object in order to conserve angular momentum. The dynamics and physics governing the flow of matter accreting onto compact objects is however still debated. BH and NS compact binary systems are powerful and highly variable sources in X-rays, and it is by studying the fluctuations in X-ray luminosities that we have made progress towards understanding the physics involved within accretion disks (\citealt{belloni_proc,vanderklis_proc}). For example, thanks to the Rossi X-Ray Timing Explorer (\textit{RXTE}), \cite{vanderklis96} have been able to detect fast (kHz) quasi-periodic oscillations (QPOs) in NS binaries, associating these with the inner edges of the accretion disk. The detected kHz QPOs can immediately be used to constrain the mass and radius of the NS (\citealt{miller}).

Variability is a common characteristic of X-ray binary systems (XRBs, consisting of NS or BH accretor), cataclysmic variables (CVs, consisting of a WD accretor) and Active Galactic Nuclei (AGN, where the accretor is a supermassive black hole). Specifically, the power spectrum of this variability consists of two components: single or multiple quasi-periodic oscillations(QPOs), and an aperiodic broad-band noise component spanning several decades in temporal frequency. QPOs have been extensively studied in a number of accreting sources, ranging from WD binaries (\citealt{warner_qpo,pretorius_qpo}), to super-massive black holes at the centre of AGN (\citealt{gierlinski}). On the other, hand the properties of the broad-band noise component has been examined in detail only in XRBs and AGN. 

One feature of the broad-band noise X-ray variability which seems to be ubiquitous in XRBs and AGN is the so-called rms-flux relation, which is a linear relationship between the absolute root-mean-square (rms) amplitude of variability and the flux, such that the sources become more variable as they get brighter (\citealt{uttley1,uttley2,gleiss,heilULX}). The rms-flux relation is remarkably linear and applies over a broad range of time-scales, so that the rms of high frequency variations correlates with those at low frequencies (\citealt{uttley2}). Crucially, the rms-flux relation is observed while the power-spectrum shape is itself stationary, implying that the rms-flux relation is a fundamental property of the variability process, and not connected with other processes which can change the variability properties on longer time-scales.  Finally, it can be shown that the rms-flux relation predicts a log-normal distribution of fluxes and implies that the variability process is non-linear (\citealt{uttley2}). The current most-favoured model to explain the rms-flux relation, as well as many other features of the variability in XRBs and AGN, is the propagating fluctuations model (\citealt{lyub,kotov,arevalo}). In this model, the observed variability is associated with modulations in the effective viscosity of the accretion flow at different radii. More specifically, the model assumes that, at each radius, the viscosity -- and hence accretion rate -- fluctuates on a local viscous time scale around the mean accretion rate, whose value is set by what is passed inwards (again on the viscous time scale) from larger radii. The overall variability of the accretion rate in the innermost disk regions is therefore effectively the product of all the fluctuations produced at all radii.

For CVs, the presence of broad-band noise in the power spectrum has been recognised for many years and is usually referred to as ``flickering'' in the literature. However, the methods employed to study XRB lightcurves have not been applied to CV data. The properties of flickering have been mainly studied through the use of eclipse maps (eg. \cite{baptista04}), which have identified the existence of two different and independent sources of flickering: the low and high-frequency flickering components. The low-frequency component originates from the overflowing gas stream from the secondary star, and is associated with the mass transfer process. The high-frequency component, on the other hand, originates from the inner accretion disk region, and no evidence of variable emission from the hot spot, gas stream, or white dwarf has been observed (contrary to some previous suggestions, see \citealt{bruch1,bruch2}). 

XRBs and AGN, which do show the rms-flux relation, share many properties with CVs. Both XRBs and CVs are observed to exhibit large-amplitude outbursts, which are theoretically explained as arising due to thermal/viscous instabilities in the accretion disks surrounding the compact objects (\citealt{ss_73}). Furthermore, similarly to XRBs, CVs also display outflowing jets in conjunction with specific changes in the spectral energy distribution (\citealt{koerding}). Therefore, it is natural to suspect that CVs should also show behaviour associated with a fluctuating accretion flow, including the rms-flux relation which is seen in accreting BH and NS. However, although CVs share many similarities with XRBs and AGN, they also differ substantially in other respects. Specifically, the radius of a WD is over an order of magnitude larger than that of a NS or the event horizon of a BH. Because of this, accreting white dwarfs are not subject to the effects of strong gravity. Additionally, accreting WDs have not been observed to display a comptonised component as in XRBs and AGN, and furthermore they do not posses the extreme magnetic fields which are found in some NS.  Therefore it remains an important open question as to whether accreting WDs should show the same rms-flux behaviour as seen in XRBs and AGN.

The broad-band noise component, present in XRBs and AGN, is well studied because of the high throughput X-ray missions with good timing capability (\textit{RXTE}, \textit{XMM-Newton}), allowing very short cadence (down to $\approx$ms) observations, at high sensitivities. In the optical, \cite{gandhi} observed three XRBs with \textit{ULTRACAM} (\citealt{ultracam}) and, as expected, also found that the lightcurves display similar rms-flux relations as those observed in the X-rays (although the optical emission could possibly originate from the jet and not the disk). Accreting white dwarfs are faint in X-rays, since their gravitational potential wells are not as deep as those of XRBs (so they show lower inner disc temperatures) and also (unlike AGN and XRBs) they do not seem to possess strong Comptonising coronae. Because of this, detailed timing studies in the X-ray region have not been possible for disk-fed white dwarfs, and most timing studies of these systems concentrated on optical light. Specifically, a series of papers (\citealt{warner_qpo,pretorius_qpo}) already noted analogies between the QPOs observed in CVs and those observed in XRBs. However, detailed analysis of the broad-band noise component in CVs was so far not as straight forward as for XRBs. The reason for this is twofold: first, the broad-band noise component displays, in the optical, low-amplitude, few per cent fractional rms variations requiring sensitive cameras; second, the broad-band noise component can only be studied with long, continuous, observations. This is because optical broad-band noise is thought to be generated further out in the accretion disk compared to X-rays, and is thus observed on longer timescales. Furthermore, broad-band noise is a stochastic process, requiring statistical studies of many realisations. This, in turn, demands long continuous observing. In order to truly assess whether accreting white dwarfs display the rms-flux relation, and thus determine if this phenomenon is indeed ubiquitous in all accreting compact objects, irrespective of accretor type, a sensitive, long, high-cadence and continuous light curve in the optical light is required. This kind of observation has only now become possible, thanks to the \Kepler\ satellite (\citealt{kepler}).

Here we analyse the \Kepler\ light curve of the accreting white dwarf system MV Lyrae (one of about 14 CVs known to exist within the \Kepler\ field-of-view), and show that it does indeed exhibit an rms-flux relation similar to that observed in Cyg X-1 by \cite{gleiss}. MV Lyrae is a VY Scl nova-like system, spending most of its time in the high state ($V\approx12-13$) and occasionally (every few years) undergoing short-duration (weeks/months) drops in brightness ($V\approx16-18$, \citealt{hoard}). It is not clear why MV Lyrae, and all VY Scl type stars, undergo this kind of behaviour, but it is known that these systems have extremely low mass transfer rates at their minimum brightness ($6\times10^{-10}M_{\sun}$ \citealt{linnell,hoard}). The inferred distance to MV Lyrae is $\approx530$pc, with a secondary star cooler than $3500$K (\citealt{hoard}). The orbital period of this CV is 3.19 hours, with a binary mass ratio of 0.43 (\citealt{skillman}). We observed MV Lyrae during it’s final ascent, in a low-to-high luminosity transition, where it’s optical light mostly originates from its nearly face-on accretion disk (\citealt{skillman}).  In the next section we will describe the data acquisition and analysis, whilst our results and discussion are presented in Section \ref{sec:discussion}.

\section{Data analysis}\label{sec:data}
The MV Lyrae light curve was obtained with the \Kepler\ satellite and provided to us by the Science Operations Centre\footnote{The data is propritary. The first five quarters of \Kepler\ observations will become public on January 7 2012} in reduced and calibrated form after being run through the standard data reduction pipeline (\citealt{jenkins}). The response function for the photometry covers the wavelength range $4000-9000$\AA. Here, we only consider the Single Aperture Photometry (SAP) light curve and do not attempt to analyse the Pre-search Data Conditioning (PDC) light curve, as the latter is optimised for planet transit searches. As a consequence of this choice, data artifacts which would have been removed by the PDC, remain in our \Kepler\ SAP light curve. These artifacts are the result of rapid attitude tweaks to correct pointing drift, caused by the loss of fine pointing accuracy. This pointing accuracy loss is due to the spacecraft thrusters firing regularly every 3 days to desaturate the reaction wheels and can affect \Kepler\ data for several minutes. Furthermore, data gaps occasionally occur due to \Kepler\ entering anomalous safe modes. During such events no data is recorded and following data is often correlated with electronic warming. Further details of artifacts within \Kepler\ light curves can be found in the \Kepler\ Data Release Notes 3\footnote{http://archive.stsci.edu/kepler/documents.html}. Here, we make no attempt to correct these artifacts, but simply remove them from the light curve. 

\begin{figure*}
\includegraphics[width=\textwidth, height=0.40\textheight]{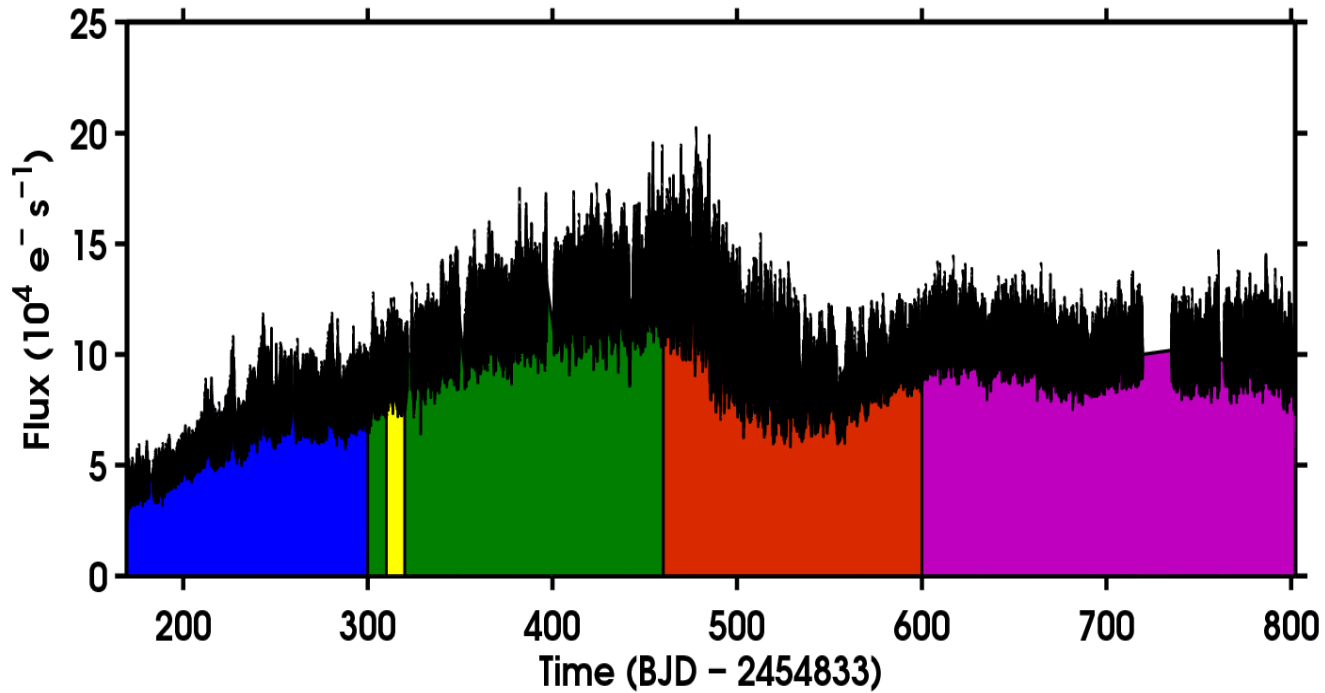}	
\caption{MV Lyrae light curve obtained by the \Kepler\ satellite. The source was observed in short cadence mode (58.8 second cadence) for 633 consecutive days. The light curve segment within the yellow shaded region was used to produce the rms vs. flux plot in Fig.\ref{fig:2}. The blue, red, green and magenta regions have been used to produce the PSDs and rms-flux relations in Fig.\ref{fig:3}, Fig.\ref{fig:4} and Fig.\ref{fig:5}. The last 233 days, marked by the magenta region, has been used to produce the flux distribution in Fig.\ref{fig:6}.}
\label{fig:1}
\end{figure*}

Fig.\ref{fig:1} shows the short cadence (58.8 seconds; \citealt{gilliland}), barycentre corrected, light curve for MV Lyrae obtained during the first four quarters of \Kepler\ operations. The lightcurve spans an interval of 633 days. The visible data gaps present in the light curve are due to the artifacts described above as well as the monthly data down-links. At first glance the light curve reveals an obvious steady increase in flux (over 1 magnitude) and also an increase in the flux variability (rms) during the rise. 

\begin{figure}
\includegraphics[width=0.45\textwidth, height=0.30\textheight]{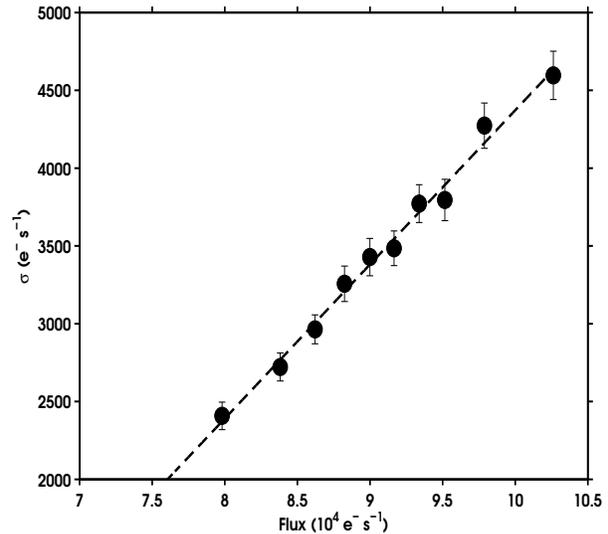}	
\caption{Rms-flux relation obtained from a 10 day segment (indicated by the yellow shade in Fig.\ref{fig:1}) in MV Lyrae.}
\label{fig:2}
\end{figure}

To characterise the relationship between the rms variability and the mean flux, we followed a similar procedure as used by \cite{gleiss} on Cyg X-1. Whilst \cite{gleiss} used the mean PSD level to measure rms, we directly measure it from the spread in the lightcurve datapoints. First, the MV Lyrae lightcurve was split into 10 day segments. For each segment we computed 10 minute flux averages and corresponding rms. All 10 day segments analysed displayed linear rms-flux relations after binning the 10 minute averages by flux. One example is shown in Fig.\ref{fig:2}. We carried out the same exercise employing smaller and larger data segments, and longer and shorter averaging timescales. In each case, we still recovered linear rms-flux relations. This is already strong evidence that the optical variability observed in the accreting WD is analogous to the X-ray variability seen in XRBs and AGN.

\begin{figure}
\includegraphics[width=0.5\textwidth, height=0.30\textheight]{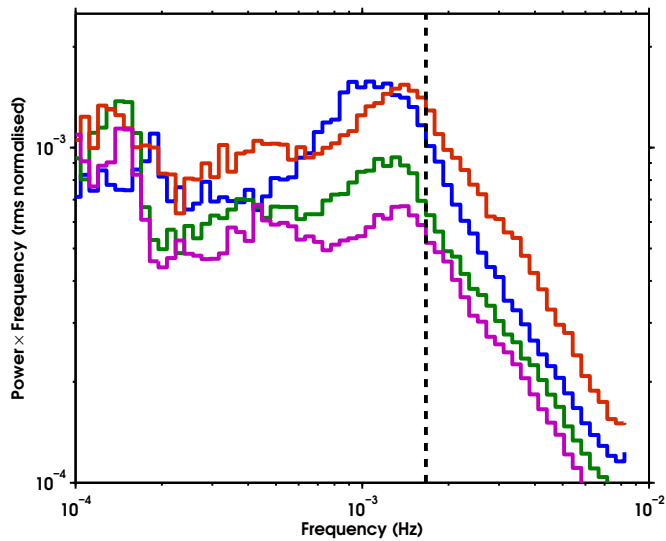}	
\caption{Power spectral distribution obtained from the MV Lyrae lighcurve shown in Fig.\ref{fig:1}. The colour coding is analogous to that in Fig.\ref{fig:1}, so that each independent PSD was obtained by the corresponding light curve segment. The Poisson noise level, ranging between $2-3\times10^{-5}$, has been subtracted. In this paper we analyse frequencies higher than those marked by the black dashed line.}
\label{fig:3}
\end{figure}

In order to quantify the linear relationship in Fig. \ref{fig:2}, we modelled it with a function of the form (\citealt{uttley1}),
\begin{equation}
\sigma=k(F-C)
\label{eq:1}
\end{equation}
where $\sigma$ is the rms value, $F$ is the flux in $\unit{e^-} \unit{s^{-1}}$, whilst $k$ and $C$ are the fit parameters. This linear model fits the general shape of the majority of our flux-sorted segments reasonably well, although occasionally large $\chi^2$ values are recovered, implying there exists some intrinsic scatter. This scatter can be caused by any non-stationarity in the variability processes which causes variations in the rms-flux relation on the 10 day time-scale of the segments. Furthermore, the presence of any frequency-varying QPO could break the linear rms-flux relation by varying the power spectrum over the frequency-range used to measure the rms. In Fig.\ref{fig:3} we show the noise-subtracted power spectral density (PSD) of the four coloured, independent, sections shown in the lightcurve of Fig.\ref{fig:1}. The timescales used to measure the rms are those at frequencies higher than 10 minutes, marked the dashed black line in Fig.\ref{fig:3}. For each segment, these rms values will respond to flux variations on longer time-scales, down to the 10 day length of the segment.

The PSDs shown in Fig.\ref{fig:3} show possible candidate QPOs at low frequencies and breaks just above $10^{-3}$Hz. From the modelling performed by \cite{arevalo}, these breaks can be associated with the viscous timescale at the inner-most edges of the accretion disk. Above the break frequency, no annuli contribute significant variability power, and the PSD bends downwards. Given that the four different quadrants appear to display shifts in the break frequency, it is possible that mass accretion rate variations are displacing the inner edges of the accretion disk. Furthermore, at frequencies below the break, there appears to be candidate QPOs. These features at low frequencies, and the PSD break shifts, will be analysed and discussed in a paper in preparation: here we are only concerned with the high-frequency component observed in MV Lyrae.

\begin{figure*}
\includegraphics[width=0.45\textwidth, height=0.30\textheight]{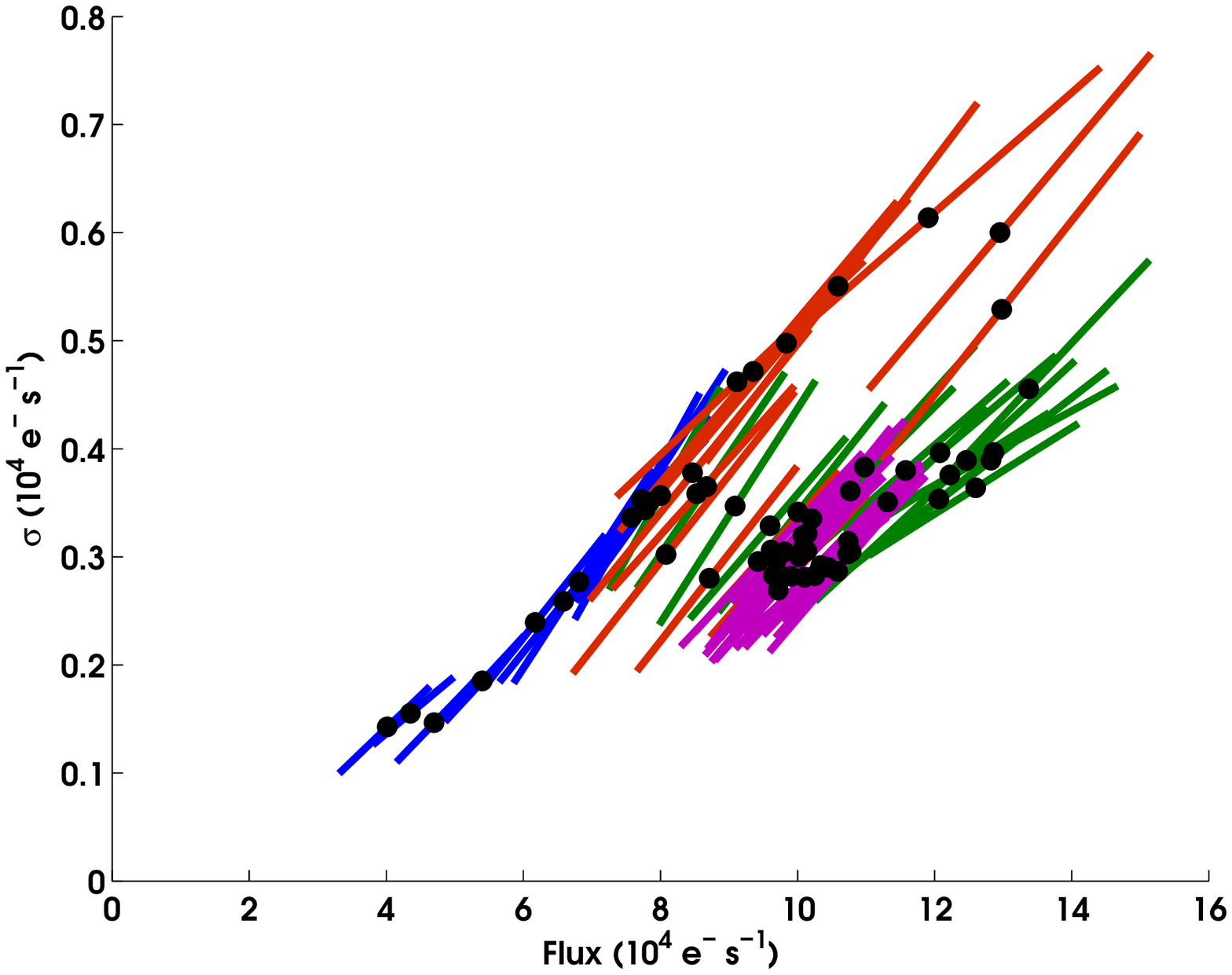}	
\includegraphics[width=0.45\textwidth, height=0.30\textheight]{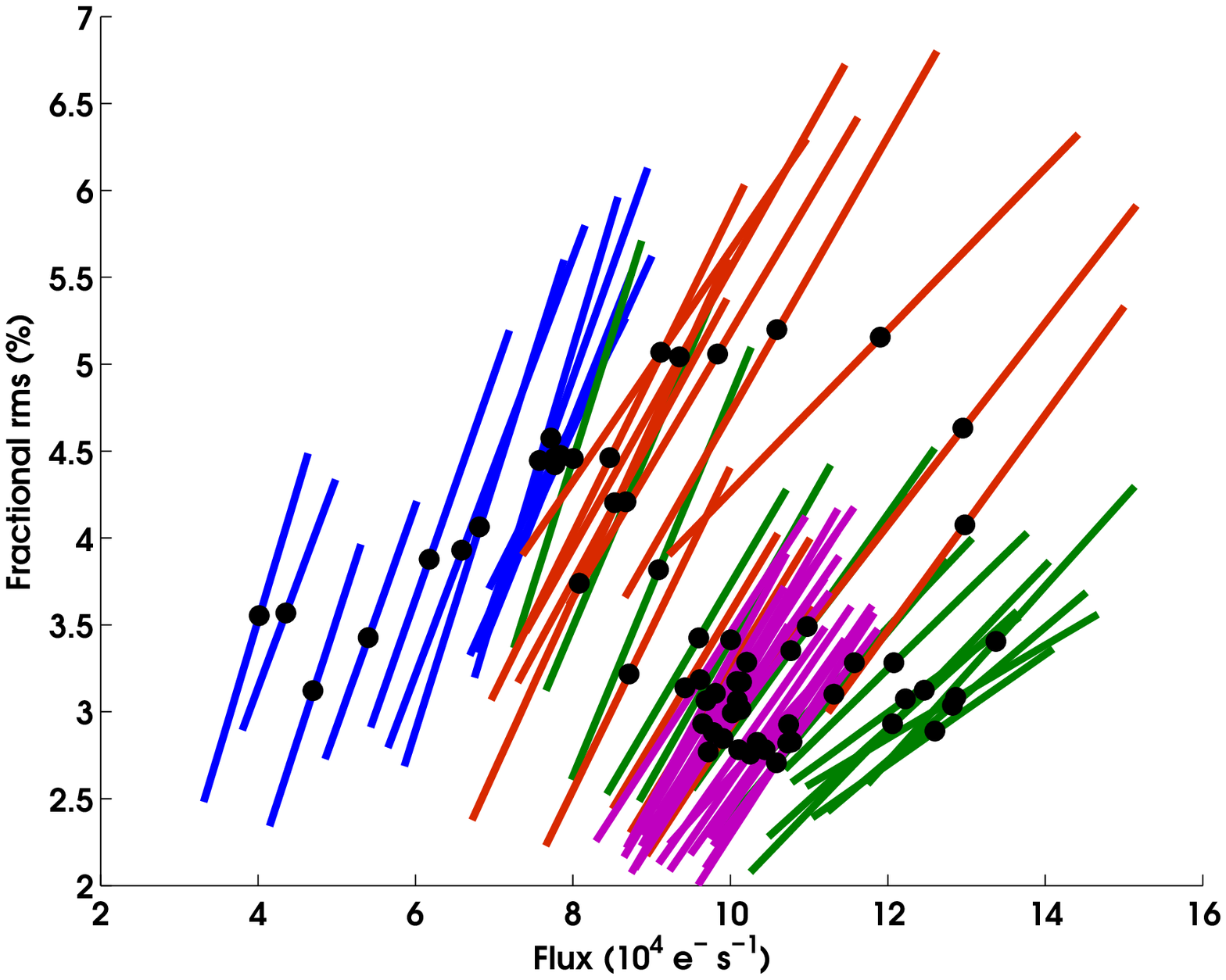}
\caption{Right: The best-fit lines for the rms-flux relations in all of our 10-day segments, where one is shown in Fig. \ref{fig:2}. In order to follow the temporal evolution of the rms-flux relations, colours are matched to those in Fig.\ref{fig:1}. The black data points are the mean flux and mean rms values of the computed rms-flux relation fits. Left: Same figure, where the y-axis displays percentage fractional rms.}
\label{fig:4}
\end{figure*}

\begin{figure}
\includegraphics[width=0.45\textwidth, height=0.30\textheight]{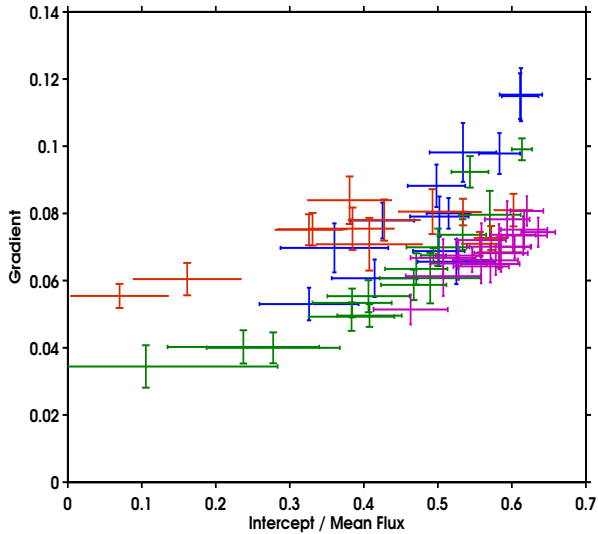}	
\caption{Best-fit parameters (gradient and intercept from Eq.\ref{eq:1} normalised by the mean flux of each 10-day segment) from the rms-flux relations shown in Fig.\ref{fig:4}. The colour coding is matched to those in Fig.\ref{fig:1}.}
\label{fig:5}
\end{figure}

Another interesting feature observed in MV Lyrae concerns the evolution of the rms-flux relation with both flux and time. In Fig.\ref{fig:4} we show the best-fit lines for the rms-flux relations in all of our 10-day segments. In order to give a feel for the temporal evolution of the rms-flux relation, each colour represents a different section in the lightcurve, with the colours in Fig.\ref{fig:1} associated to those in Fig.\ref{fig:4}. Specifically, there appears to be a hysteresis-like behaviour in the evolution of the rms-flux relation within the broad-band noise of MV Lyrae. After about 150 days, the rms-flux relations seem to change behaviour and detach from the original trend (marked with blue lines in Fig.\ref{fig:4}) and move to higher fluxes maintaining similar rms variability. In proximity of the peak flux in Fig.\ref{fig:1}, after about 250 days of observations, the rms-flux relations seem to change behaviour once again during their flux descent (red lines), seemingly looping back to their original track after having reached a maximum rms variability of about $6\%$. This behaviour has not been observed in any other accreting compact binary or AGN, and it is not clear at the time of writing what the physics behind this phenomenon is. In Fig.\ref{fig:5} we show the gradient and intercept for the best-fit lines in Fig.\ref{fig:4}, colour coded in the same way. Similarly to the results obtained on the BH binary Cyg X-1 by \cite{gleiss}, the gradient and intercept in MV Lyrae seem to be correlated as well. This suggests that some fundamental and analogous process is giving rise to the rms-flux behaviour in both XRBs and CVs. Interesting is also the fact that in the gradient vs. intercept plane there seem to be two distinct ``tracks'': the rms-flux relations obtained from the blue and red lightcurve segments seem to have a systematic offset in the gradient when compared to the green and magenta segments. Although we cannot properly quantify this behaviour at the moment with the available data, the evolution in Fig.\ref{fig:4} also seems to suggest that the rms-flux behaviour systematically changes during the lightcurve evolution.

Finally, we also examined the flux distribution within the MV Lyrae lightcurve. Specifically, the observed fluxes in XRBs and AGN which do show the rms-flux relation are log-normally distributed. In order to perform this test, we produced a histogram of fluxes from the last 233 days (marked by the magenta region in Fig.\ref{fig:1}) of the MV Lyrae lightcurve. We then fitted the histogram with both a symmetric Gaussian and a three-parameter log-normal model. The number of flux measurements per bin ($N$) was required to be at least 50 for the fits, with bin error assigned as $\sqrt{N}$. We performed this exercise on the last 233 days of the lightcurve, as these appear to be most the stationary in mean flux: a requirement for the log-normal behaviour to be expected (see \citealt{uttley2,gandhi}). Statistically, the Gaussian fit resulted in the goodness-of-fit parameter $\chi^2$/d.o.f of 18802/269, whilst the log-normal fit resulted in 334/269. MV Lyrae clearly exhibits a log-normal flux distribution. The excess residual of our fit can be explained by the weak non-stationarity of the lightcurve, as also explained in \cite{uttley2} and \cite{gandhi}.

\begin{figure}
\includegraphics[width=0.45\textwidth, height=0.30\textheight]{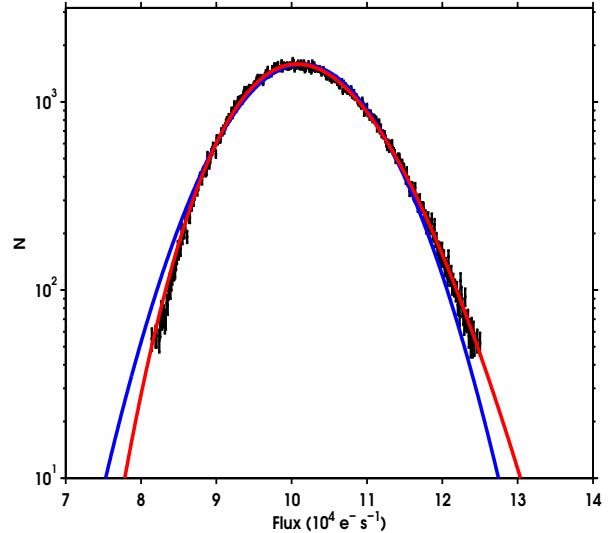}	
\caption{Binned frequency distribution of the mean source flux for the magenta section of the MV Lyrae light curve in Fig.\ref{fig:1} (black data points). A log-normal distribution fit is shown in red and a Gaussian fit in blue. A log-normal fit statistically describes the data better than a Gaussian.}
\label{fig:6}
\end{figure}

\section{Discussion}\label{sec:discussion}
We showed how the accreting WD MV Lyrae obeys a linear rms-flux relation like many other BH/NS compact interacting binaries and AGN. This is the first time such a correlation was found in an accreting object other than a BH/NS X-ray binary or AGN. Furthermore, we showed how the rms-flux relation evolves with time, displaying, what appears to be, a hysteresis-like behaviour. The physics of this phenomenon is not yet understood, and it is the first time it has been observed in an accreting compact object. Additionally, we confirmed our findings by ascertaining that the flux distribution of MV Lyrae follows a log-normal distribution, as seen in other accreting compact objects displaying the linear rms-flux relation.

The leading interpretation of the rms-flux relation is that of a fluctuating accretion disk (\citealt{lyub}), where variations in the accretion rate occur at all radii and then propagate inwards on the viscous timescale within the accretion disk. These fluctuations couple together in a multiplicative way as they propagate down to the centre, so that low frequency fluctuations produced at large radii modulate the higher frequency fluctuations produced further in. The multiplicative nature of the process will yield flux variations which follow a log-normal distribution, like the one in Fig.\ref{fig:6}. 

\cite{chur} have pointed out that, in a standard thin disk, viscous damping will smooth accretion rate fluctuations on the viscous timescale. This means that, for the \cite{lyub} model to work, the fluctuations must happen in a geometrically thick accretion disk flow (such as an advection dominated flow or thick disk). We can use the PSD break just above $10^{-3}$Hz ($\approx13$ minutes, shown in Fig.\ref{fig:3}), or the relative rms variations, to constrain the viscosity parameter, $\alpha$, and the disk scale height, $H/R$.

\begin{figure}
\includegraphics[width=0.45\textwidth, height=0.30\textheight]{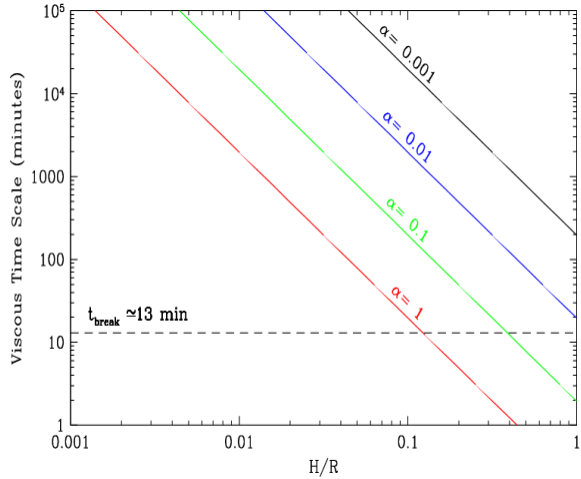}	
\caption{Viscous timescale as a function of $H/R$ from the standard $\alpha$-disk solution (\citealt{FKR}, Eq.~5.68). Each colour corresponds to a different $\alpha$ with black, blue, green and red corresponding to 0.001, 0.01, 0.1 and 1 respectively. The horizontal line marks the PSD break at $\approx13$ minutes.}
\label{fig:7}
\end{figure}

The PSD break in Fig.\ref{fig:3} can be associated to the viscous timescale at the inner edges of the accretion disk (\citealt{arevalo}). If this were the case, we would require a geometrically thick disk of about $H/R \approx 0.1$ (where $H/R$ is the ratio of the vertical scale height to radius in the disk), and a viscosity parameter $\alpha \approx 1$. This is shown in Fig.\ref{fig:7}, where we computed the viscous timescales resulting from the standard $\alpha$-disk solution of \cite{FKR} (Eq.~5.68) as a function of $H/R$ for an accretion disk extending all the way down to the WD surface at $7\times10^{8}$cm. Both these values are considered to be high for a standard disk solution, and are pushing into uncomfortable limits of the \cite{ss_73} parameter space.

Following the procedure adopted by \cite{baptista04} on V2051 Oph, we also used the rms variability amplitude to obtain $\alpha$ and $H/R$. From the inferred $3\%$ rms variability (see Fig.\ref{fig:4}) we obtain, similarly to V2051 Oph, $\alpha\approx0.16$. This estimate assumes an $H/R$ value of about $10^{-2}$, and is based on the effects of magneto-hydrodynamic (MHD) turbulence in an accretion disk, as modelled by \cite{geertsema}. However, when the variability reaches $6\%$ in MV Lyrae, $\alpha\approx0.62$, again pushing into uncomfortable limits of the \cite{ss_73} parameter space. To make $\alpha$ consistent with the expected viscosity of hot disks (\citealt{lasota01}), the disk scale height, $H/R$, would have to increase to $0.1$. This would then be in line with a geometrically thick, hot, disk. If this were the case, the rms variability amplitude would be an indicator of the disk scale height (assuming $\alpha$ to remain constant at it's hot value of $\approx0.1$ -- $0.2$). More importantly, all the above estimates for the disk scale height and $\alpha$ are suggesting that the propagating fluctuation models may have difficulty in explaining the variability if the disc behaves like a ``standard'' thin disc. One way to overcome these inconsistencies would be to invoke an overflowing gas stream, as mentioned by \cite{baptista04}. This flow could exist above the standard thin accretion disk, extending all the way close to the WD surface, and possibly provide the thick disk required to explain the fast viscous timescale inferred from the PSD breaks.

One other important implication of our discovery regards the interpretation of flickering noise in CVs in general. It has been recognised for many decades that almost all CVs display at least some optical flickering. Many suggestions have emerged in order to explain the origin of this flickering, by associating it to the accretion hot spot or the boundary layer close to the WD surface (\citealt{bruch1,bruch2}). More recently, \cite{baptista04} showed that CV flickering is caused by the existence of two different and independent sources. The low-frequency flickering component is caused by the overflowing gas stream from the secondary star, whilst the high-frequency component by the inner-most regions of the accretion disk. Our result is in agreement with the interpretation that the high-frequency flickering is created within the accretion disk. The fact that we observe the linear rms-flux relation within MV Lyrae (a canonical flickering CV) gives us confidence that the fluctuating accretion disk model is also at play within other CV systems. Furthermore, the observed fractional rms amplitude of $\approx3\%$, generally observed in MV Lyrae (see Fig.\ref{fig:4}), is also similar to that found in V2051 Oph by \cite{baptista04}.

Although it is clear from the \cite{lyub} model that the variability is the result of modulations from an ensemble of coupled perturbations, the physical mechanism of variable emission is not yet known. Some suggested that synchrotron emission at the base of the outflowing jet could be a possibility (\citealt{king}). However MV Lyrae is not known to display a jet (or synchrotron emission), and it is not likely that a comptonised component is driving the rms-flux relation out to the large radii observed in the optical. More likely instead is that instabilities within the accretion disk are responsible for driving the continuum variability. 

This interpretation is also supported by the recent results of \citealt{uttley3}. They showed that for XRBs the blackbody component in X-rays, associated with the disk, substantially leads the power law component associated with the comptonised region close to the compact object. Thus, the comptonised region, thought to drive jet-launching, is responding to changes in the accretion disk, and not vice-versa. In this respect, we would expect MV Lyrae to display energy-dependent time lags in optical light (\cite{arevalo}), where the broad-band variability arising from bluer bands is lagging the variability observed from the redder ones: simply a consequence of the fact that the fluctuations are propagating inwards within the disk, affecting the redder outer edges before the bluer inner ones. 

The rms-flux relation remains an enigmatic observational feature of all accreting compact objects, but it clearly contains information about the dynamics of the infalling material. Although the microphysics of accretion disk dynamics are not yet fully understood, our result strongly suggests that the processes involved are the same in all accreting compact objects, independent of mass, type and size. Our result has thus revealed yet another connection between accreting compact objects, suggesting that the physics of disk accretion are universal.

\section*{Acknowledgements}
This paper includes data collected by the \Kepler\ mission. Funding for the \Kepler\ mission is provided by the NASA Science Mission directorate. This research has made use of NASA's Astrophysics Data System Bibliographic Services. S.S. acknowledges funding from NWO project 600.065.140.08N306 to P.J. Groot. P.U. acknowledges funding from an STFC Advanced Fellowship and from the European Community's Seventh Framework Programme (FP7/2007-2013) under grant agreement number ITN 215212 ``Black Hole Universe''. M.S. acknowledges funding from the NASA grant NNX11AB86G.

\bibliographystyle{mn2e}
\bibliography{MVLyr_paper}

\label{lastpage}

\end{document}